\documentclass[conference]{IEEEtran}
\IEEEoverridecommandlockouts

\usepackage{url}
\usepackage[colorlinks,allcolors=blue]{hyperref}

\usepackage[numbers,sort]{natbib}

\usepackage{amsmath,amssymb,amsfonts,mathtools}
\usepackage{graphicx}
\usepackage{textcomp}
\usepackage[hyphenbreaks]{breakurl}
\usepackage{xcolor}

\usepackage{algorithm}
\usepackage{algpseudocode}

\usepackage{booktabs}
\usepackage{adjustbox}

\usepackage{tikz}
\usepackage{pgfplots}
\usepgfplotslibrary{groupplots}
\pgfplotsset{compat=newest}

\def \given {\,|\,}
\def \d {\mathrm{d}}
\def \kl {\mathrm{KL}}
\def \N {\mathrm{N}}

\newcommand{\bP}{\bar{P}}
\newcommand{\bm}{\bar{m}}

\def\BibTeX{{\rm B\kern-.05em{\sc i\kern-.025em b}\kern-.08em
		T\kern-.1667em\lower.7ex\hbox{E}\kern-.125emX}}

\begin{document}

\title{Variational Gaussian filtering \\ via Wasserstein gradient flows\\
	\thanks{Adrien Corenflos is funded by the Academy of Finland project 321891 (ADAFUME). Hany Abdulsamad is funded by the Finnish Center for Artificial Intelligence (FCAI).}
}

\author{
	Adrien Corenflos\IEEEauthorrefmark{1}, Hany Abdulsamad\IEEEauthorrefmark{2} \\
	\textit{Department of Electrical Engineering and Automation} \\
	\textit{Aalto University, Finland} \\
	Email: \IEEEauthorrefmark{1}{adrien.corenflos@aalto.fi}, \IEEEauthorrefmark{2}{hany.abulsamad@aalto.fi}
}

\maketitle

\begin{abstract}
	We present a novel approach to approximate Gaussian and mixture-of-Gaussians filtering. Our method relies on a variational approximation via a gradient-flow representation. The gradient flow is derived from a Kullback--Leibler discrepancy minimization on the space of probability distributions equipped with the Wasserstein metric. We outline the general method and show its competitiveness in posterior representation and parameter estimation on two state-space models for which Gaussian approximations typically fail: systems with multiplicative noise and multi-modal state distributions.
\end{abstract}


\section{Introduction}\label{sec:intro}
State-space models (or hidden Markov models) are a class of models widely used to represent latent dynamics that are partially or indirectly observed. They typically arise in ecological, economical, and tracking applications~\cite[for an introduction, see, e.g.][]{jazwinski2007stochastic}. Formally, state-space models are given by a set of dynamics and noisy observations, often depending on a set of parameters $\theta$
\begin{equation}\label{eq:ssm}
	\begin{aligned}
		X_0 & \sim p_0(\cdot \given \theta), \\
		X_{k+1} & \sim p_k(\cdot \given X_k, \theta),\\
		Y_k & \sim h_k(\cdot \given X_k, \theta).
	\end{aligned}
\end{equation}
While the problem of inference in such models is generally intractable, computing the filtering distribution $p(x_k \given y_{0:k})$ can typically be done exactly if the state-space is finite ($x_k$ can only take a finite number of values) or when all the (conditional) densities in \eqref{eq:ssm} are Gaussian using the celebrated Kalman filter~\cite{Kalman1960}. When this is not the case, relying on approximations becomes necessary. Two important types of approximations are approximate Gaussian filters~\cite[see, e.g.][]{sarkka2013bayesian}, and Monte Carlo filters~\cite[see, e.g.][]{Chopin2020book}.

Although the standard filtering problem is important, one may also be interested in system identification, which, in the parametric context, refers to learning $\theta$ from a sequence of observations $\{y_{0}, \ldots, y_{K}\}$.

In this article, we pay particular attention to two classes of models, typically neglected in approximate Gaussian filtering. The first class is that of models with multiplicative noise, for which stochastic volatility models are an illustrative example, often used in economics to model financial returns~\cite{heston1993closed}. These are usually given as an auto-regressive latent state $x_k$, and observations $y_k$ following
\begin{equation}\label{eq:stochvol}
	\begin{aligned}
		Y_k & = \exp(X_k/2) \, \eta_k, \\
		X_{k+1} & = \mu + \alpha(X_{k} - \mu) + \sigma \, \epsilon_{k},
	\end{aligned}
\end{equation}
where the noise processes are correlated
\begin{equation}\label{eq:corr-noise}
	\begin{pmatrix}
		\epsilon_k \\ \eta_k
	\end{pmatrix} 
	\sim \N\left(
	\begin{pmatrix}
		0 \\ 0
	\end{pmatrix},
	\begin{pmatrix}
		1 & \rho \\ \rho & 1
	\end{pmatrix}
	\right).
\end{equation}
The second class we consider are systems for which the state (filtering) posterior is a multi-modal distribution. A simple example of this form can be given by constructing a latent state $x_k$ with random walk dynamics while the observations $y_k$ are a function of  the absolute value of $x_k$:
\begin{equation}\label{eq:multimodal}
	\begin{aligned}
		X_{k+1} & = X_{k} + \epsilon_k, && \, \epsilon_k \sim \N(0, 1), \\ 
		Y_k & = |X_k| + \eta_k, && \, \eta_k \sim \N(0, 1).
	\end{aligned}
\end{equation}
If, for example, $X_0 \sim \N(0, \delta^2)$, it is straightforward to see that the state filtering distribution will be bi-modal and fully symmetric with respect to the x-axis.

\subsection{Contributions}\label{subsec:contribs}
Existing approximate Gaussian filtering methods suffer from several drawbacks. The linearization methods of~\cite{garcia2015posterior, Tronarp2018iterative} for example require computing conditional expectations $m_k(x) = \mathbb{E}\left[Y_k \given X_k=x\right]$. In the case of the stochastic volatility model~\eqref{eq:stochvol}, this quantity will unequivocally be null (at least for $\rho = 0$, see Section~\ref{subsec:exp-stoch-vol} for details). Consequently, applying these methods to \eqref{eq:stochvol} will lead to a filtering (or a smoothing) solution independent of the gathered observations, which is problematic.

Additionally, classical linearization methods do not extend directly to multi-modal distributions, and they do not, to the best of our knowledge, enable handling mixtures of Gaussians. In view of this, our contributions are the following:
\begin{enumerate}
	\item We rephrase the filtering problem as an iterative distribution fitting problem.
	\item We apply the variational inference method from \cite{Lambert2022} to propagate Gaussian approximations between time steps and formulate it as a fixed point iteration for efficient gradient calculation.
	\item We use our method for parameter estimation in stochastic volatility models and filtering of multi-model target distributions.
\end{enumerate}

\section{Methodology}

\subsection{Variational Inference via Wasserstein Gradient-flows} \label{sec:vi-gradient-flow}
Let $\pi(x) \propto \exp(-V(x))$ be an arbitrary target distribution known up to a normalizing constant. Given a variational family of distributions, $q_{\phi}(x)$, $\phi \in \Phi$, and a measure of discrepancy $L(\phi) = D(\pi, q_{\phi})$ between $\pi$ and $q_{\phi}$, it is natural to try and approximate $\pi$ with $q_{\phi}$ by minimizing $L$.

A typical discrepancy is given by the Kullback--Leibler (KL) divergence  \cite{joyce2011kullback}
\begin{equation}
	L(\phi) = \kl(q_{\phi} \given \pi) \coloneqq \int q_{\phi}(x) \log \frac{q_{\phi}(x)}{\pi(x)} \, \d x.   
\end{equation}
This divergence presents a number of attractive properties for statistical inference: (i) it is positive, (ii) it only requires evaluating $V(x)$ and does not necessitate knowing the normalizing constant of $\pi$, and (iii), it is exact in the sense that $\kl(q_{\phi} \given \pi) = 0$ if and only if $q_{\phi} = \pi$. For more details, we refer the reader to~\cite{murphy2022book}.

For example, if $q_{\phi}(x) = \N(x \given \mu, \Sigma)$ is in the family of well-defined Gaussians, parameterized by their mean and covariance, then the inference problem can be cast as a minimization of $L(\cdot)$ with respect to $(\mu, \Sigma)$. Furthermore, when ignoring the positivity constraints on $\Sigma$, it is possible to define a gradient flow on $\phi = (\mu, \Sigma)$, akin to a gradient descent on $L(\cdot)$ in continuous time
\begin{equation}\label{eq:euclidean-flow}
	\frac{\d \phi_t}{\d t} = -\nabla_{\!\phi} L(\phi_t),
\end{equation}
which, under an assumption of convexity, will converge to the minimizer of the objective $L(\cdot)$~\cite{santambrogio2017gradientflows}.

While this procedure is correct in essence, it targets the problem indirectly by first assuming an arbitrary parameterization of the model which may or may not respect convexity. A more direct approach is to fit $q_{\phi}$ to $\pi$ in terms of a minimization problem over the space of probability distributions, where we want to minimize $L(q) = D(\pi, q)$. In this case, it is possible to define an analog to the gradient flow~\eqref{eq:euclidean-flow} by equipping the space of probability distributions with the Wasserstein distance~\cite[Chap. 6]{Villani2009}. Under this metric, we can define a trajectory of probability distributions $q_t(x) \in \mathcal{P}(\mathbb{R}^d)$ via a partial differential equation
\begin{equation}\label{eq:wasserstein-flow}
	\frac{\partial q_t(x)}{\partial t} = \nabla \cdot \left[ q_t(x) \nabla \log \frac{q_t(x)}{\pi(x)} \right],
\end{equation}
where $\nabla \cdot$ is the divergence operator, expressed in Euclidean coordinates. 

Interestingly, by restricting $q_t$ to represent a Gaussian distribution, it was shown in~\cite{Lambert2022}, following~\cite{sarkka2007unscented}, that~\eqref{eq:wasserstein-flow} can be reformulated into coupled ordinary differential equations (ODEs) on the mean $\mu_t$ and covariance $\Sigma_t$ of $q_t$
\begin{equation}\label{eq:mean-cov-flow}
	\begin{aligned}
		\frac{\d \mu_{t}}{\d t} & = -\mathbb{E} \left[\nabla V(Z_t)\right] \\
		\frac{\d \Sigma_t}{\d t} & =
		\begin{aligned}[t]
			& 2 \, I - \mathbb{E} \left[\nabla V(Z_t) \otimes (Z_t - \mu_t) \right] \\
			& - \mathbb{E} \left[(Z_t - \mu_t) \otimes \nabla V (Z_t)\right],
		\end{aligned}
	\end{aligned}
\end{equation}
where $I$ is the identity matrix with dimensions $d \times d$ and $Z_{t} \sim \N(\mu_t, \Sigma_t)$ is Gaussian. Now, provided that we can compute or approximate the expectations arising in~\eqref{eq:mean-cov-flow}, we can find a minimizer $q^{*}(x) = \N(x \given m, P)$ of $L(\cdot)$ by integrating these coupled ODEs until convergence.

\subsection{Filtering as Variational Inference}\label{subsec:filtering-vi}
The filtering problem involves inferring the posterior distribution $p(x_k \given y_{0:k})$ for each time step $k$. To do so, it is often possible to rely on the familiar \emph{innovation-prediction} decomposition~\citep[Chap. 4]{sarkka2013bayesian}
\begin{equation}\label{eq:bayes-filtering}
	\begin{aligned}
		p(x_k \given y_{0:k}) & \propto p(y_k \given x_k) p(x_k \given y_{0:k-1}), \\
		p(x_k \given y_{0:k-1}) & = \! \int \! p(x_k \given x_{k-1}) p(x_{k-1} \given y_{0:k-1}) \, \d x_{k-1}.
	\end{aligned}
\end{equation}
For simplicity, we will focus on state-space models with affine Gaussian transition models
\begin{equation}
	p(x_k \given x_{k-1}) = \N(x_k \given A_{k-1} \, x_{k-1} + b_{k-1}, Q_{k-1}),
\end{equation}
covering those previously presented in \eqref{eq:stochvol} and \eqref{eq:multimodal}. 

In that case, assuming a Gaussian approximation is available for the previous time step $k-1$,
\begin{equation}
	p(x_{k-1} \given y_{0:k-1}) \approx \N(x_{k-1} \given m_{k-1}, P_{k-1}),
\end{equation}
then the \emph{prediction} step of the filter leads to another Gaussian distribution $p(x_k \given y_{0:k-1}) \approx \N(x_k \given \bm_{k}, \bP_{k})$ with mean and covariance
\begin{equation}
	\begin{aligned}
		\bm_k & = A_{k-1} \, m_{k-1} + b_{k-1}, \\
		\bP_k & = A_{k-1} \, P_{k-1} \, A_{k-1}^{\top} + Q_{k-1}.
	\end{aligned}
\end{equation}

Given this predictive distribution, the \emph{innovation} step computes the approximate filtering distribution 
\begin{equation}
	p(x_k \given y_{0:k}) \propto \exp(-V(x_k)), 
\end{equation}
where the potential function $V(x_k)$ is given as
\begin{equation}\label{eq:potential}
	V(x_k) = -\log p(y_k \given x_k) - \log \N(x_k \given \bm_k, \bP_k).
\end{equation}
Consequently, in order to find the parameters of a variational Gaussian $\N(x_k \given m_k, P_k)$ that approximates the posterior $p(x_k \given y_{0:k})$, it suffices to follow the recipe from Section~\ref{sec:vi-gradient-flow} and integrate~\eqref{eq:mean-cov-flow} up to stationarity, starting from the predictive parameters $(\bm_k, \bP_k)$, or any other approximation of $(m_k, P_k)$. 

When the transition $p(x_k \given x_{k-1})$ is not Gaussian, it is possible to use a similar variational approach to propagate the Gaussian approximation $p(x_{k-1} \given y_{0:k-1}) \approx \N(x_{k-1} \given m_{k-1}, P_{k-1})$ through the non-Gaussian dynamics and approximate the marginal  $p(x_{k} \given y_{0:k-1})$ with another Gaussian $\N(x_k \given \bm_k, \bP_k)$. This method was used in \cite{sarkka2007unscented} to propagate the Gaussian approximation through dynamics defined by a stochastic differential equation. Combining this with our approach is, therefore, \emph{de facto} possible. However, due to the limited scope, we leave this aspect for future work and only consider Gaussian dynamics in the remainder of this article.

Finally, because the likelihood of the observations is given by $p(y_{0:k}) = p(y_{0:k-1})\int p(y_k \given x_k) p(x_k \given y_{0:k-1}) \, \d x_k$, it is easy to derive an approximation of the marginal log-likelihood of the model by recursion. That is because the quantity $\ell_{k} = \int p(y_k \given x_k) p(x_k \given y_{0:k-1}) \, \d x_k$ is evaluated as part of \eqref{eq:mean-cov-flow}, and can therefore be used to provide an online estimate of the log-likelihood increments, which, in turn, can be used, for example, in a system identification scenario. We return to this point in Sections~\ref{subsec:details} and~\ref{subsec:exp-stoch-vol}.

\subsection{The Multi-modal Filtering Case} \label{subsec:multi-modal}
To generalize the variational technique from the previous section, let us suppose that the filtering distribution at time $k-1$ is instead given by a mixture density 
\begin{equation}
	p(x_{k-1} \given y_{0:k-1}) = \sum_{i=1}^{N_i} w^{(i)} \, \N(x_{k-1} \given m_{k-1}^{(i)}, P_{k-1}^{(i)}),
\end{equation}
with $w^{(i)} = 1/N_i, \forall i \in [1, N_i]$. In this case, when the transition dynamics are Gaussian, it is easy to show that 
\begin{equation}\label{eq:pred-mixture}
	p(x_{k} \given y_{0:k-1}) = \sum_{i=1}^{N_i} w^{(i)} \, \N(x_k \given \bm_{k}^{(i)}, \bP_{k}^{(i)}),
\end{equation}
where $\forall i \in [1, N_i]$
\begin{equation}
	\begin{aligned}
		\bm_k^{(i)} & = A_{k-1} \, m_{k-1}^{(i)} + b_{k-1}, \\
		\bP_k^{(i)} & = A_{k-1} \, P_{k-1}^{(i)} \, A_{k-1}^{\top} + Q_{k-1}.
	\end{aligned}
\end{equation}
These updates correspond to a tractable \emph{prediction} step. As a consequence, we only need to understand how to perform the $\emph{innovation}$ step to arrive at the posterior $p(x_{k} \given y_{0:k})$.

Conveniently, \cite{Lambert2022} also shows that the duality between the gradient flow~\eqref{eq:wasserstein-flow} and the coupled ODEs~\eqref{eq:mean-cov-flow} extends to the case of the finite variational mixture of Gaussians
\begin{equation}
	q_t(x) = \sum_{i=1}^{N_i} \omega^{(i)} q_t^{(i)}(x)= \sum_{i=1}^{N_i} \omega^{(i)} \, \N(x \given \mu_t^{(i)}, \Sigma_t^{(i)}),
\end{equation}
with $\omega^{(i)} = 1/N_i, \forall i \in [1, N_i]$. In this case, rather than a pair of ODEs, we obtain a system of such ODEs
\begin{equation} \label{eq:mixture-mean-cov-flow}
	\begin{aligned}
		\frac{\d \mu_t^{(i)}}{\d t} & = - \mathbb{E} \left[\nabla \log \frac{q_t}{\pi}(Z_t^{(i)}) \right] \\
		\frac{\d \Sigma_t^{(i)}}{\d t} & = 
		\begin{aligned}[t]
			& - \mathbb{E} \left[\nabla^2 \log \frac{q_t}{\pi}(Z_t^{(i)})\right] \Sigma_t^{(i)} \\
			& - \Sigma_t^{(i)} \mathbb{E} \left[\nabla^2 \log \frac{q_t}{\pi}(Z_t^{(i)}) \right],
		\end{aligned}
	\end{aligned}
\end{equation}
where, for all $i$, $Z_t^{(i)} \sim \N(\mu_t^{(i)}, \Sigma_t^{(i)})$ is Gaussian, and $\nabla^2$ denotes the Hessian operator. 

This result means that, given a Gaussian mixture approximation of the predictive $p(x_k \given y_{1:k-1})$, we can obtain a Gaussian mixture approximation of $p(x_k \given y_{0:k})$ by integrating \eqref{eq:mixture-mean-cov-flow} and following a similar approach to Section~\ref{subsec:filtering-vi}. 

\subsection{Numerical Considerations and Implementation} \label{subsec:details}
In practice, the integrals arising in~\eqref{eq:mean-cov-flow} and~\eqref{eq:mixture-mean-cov-flow} are not available in closed form. Therefore, we need to resort to numerical integration. This can be done by using any form of deterministic or stochastic Gaussian quadrature, for example, Monte Carlo~\cite[see, e.g.,][]{pages2018numerical} or sigma-points~\cite[see, e.g.,]{sarkka2013bayesian} methods. The two approaches have their pros and cons: Monte Carlo will give the correct solution on average, while deterministic quadrature is bound to be biased. Nonetheless, deterministic rules are sometimes more practical when no stochasticity should be allowed in the system. Whichever is chosen, we will obtain an approximation
\begin{equation}\label{eq:mean-cov-flow-approx}
	\frac{\d \mu_t}{\d t} \approx F_m(\mu_t, \Sigma_t), \quad \frac{\d \Sigma_t}{\d t}\approx F_P(\mu_t, \Sigma_t),
\end{equation}
of~\eqref{eq:mean-cov-flow} for the choice of $V(x_k)$ given by~\eqref{eq:potential}. Moreover, the same approximation scheme can be used to compute log-likelihood increments $\log \mathbb{E} \left[p(y_k \given x_k)\right]$ under the predictive $\N(x_k \given \bm_{k}, \bP_{k})$. In Section~\ref{sec:examples}, we use Gauss--Hermite~\cite[see, e.g.,][Chap. 5]{sarkka2013bayesian} quadrature integration rules with order five.

\begin{algorithm}[!t]
	\caption{Uni-modal Wasserstein Gradient-flow Filter}
	\label{alg:variational-filtering}
	\begin{algorithmic}[1]
		\item \textbf{input:} Measurements $y_{0:K}$ and prior $(m_0, P_0)$
		\vspace{2pt}
		\item \textbf{output:} Filtering posterior distributions $(m_{1:K}, P_{1:K})$ and marginal log-likelihood $\ell = \log p(y_{0:K})$
		\vspace{2pt}
		\State Set $\ell \gets 0$
		\For{$k \gets 1$ \textbf{to} $K$}
		\State Set $\ell_k \gets \log \mathbb{E}_{x \sim \N(\bm_k, \bP_k)}p(y_k \given x_{k})$
		\State Set $\ell \leftarrow \ell + \ell_k$ \Comment{Log-likelihood}
  		\State $(m_k, P_k) \gets (\bm_k, \bP_k)$ 
		\While{Not converged} \Comment{Innovation}
		\State $(m_k, P_k) \gets I(m_k, P_k)$ 
		\EndWhile
		\State Set $\bm_{k+1} \gets A_{k} \, m_k + b_k$ \Comment{Prediction}
		\State Set $\bP_{k+1} \gets A_{k} \, P_k \, A_{k}^{\top} + Q_{k}$
		\EndFor \\
		\Return $(m_{1:K}, P_{1:K})$, $\ell$
	\end{algorithmic}
\end{algorithm}

We now assume that we have chosen an integration method $I$ for which the stationary solution of~\eqref{eq:mean-cov-flow-approx} is a fixed point: $(m_k, P_k) = I(m_k, P_k)$. This can, for example, be an Euler integration step with a small step size. This fixed-point perspective allows us to leverage the implicit function theorem and bypass the loop when computing gradients for system identification. For the sake of brevity, we omit the details here and refer to~\cite{christianson1994fixedpoint} for the method and to our code\footnote{Implementation of the fixed-point iteration is available under \url{https://github.com/hanyas/wasserstein-flow-filter/blob/master/wasserstein_filter/utils.py}} for a Python implementation.

Therefore, the final algorithm for variational uni-modal filtering from Section~\ref{subsec:filtering-vi} is given by Algorithm~\ref{alg:variational-filtering}. Due to space constraints, we do not reproduce the algorithm for the multi-modal case from Section~\ref{subsec:multi-modal}: the procedure follows the same steps, albeit for a larger system of ODEs.

\section{Empirical Evaluation}\label{sec:examples}
As discussed in Section~\ref{sec:intro}, introducing our method is motivated by the problems posed by multiplicative noise and multi-modality in approximate Gaussian filtering. Consequently, we aim to demonstrate its effectiveness on these two problems. We compare the variational Wasserstein filter (VWF) to a bootstrap particle filter (PF, \cite{gordon1993novel}) using $500$ particles and the \emph{continuous} resampling scheme from~\cite{malik2011continuous}. This resampling allows using the particle filter in a standard parameter estimation scenario. The code to reproduce these experiments can be found at \url{https://github.com/hanyas/wasserstein-flow-filter}.

\subsection{Stochastic Volatility with Leverage}\label{subsec:exp-stoch-vol}
First, we consider the stochastic volatility model as given by~\eqref{eq:stochvol} and~\eqref{eq:corr-noise}. 
Because the noises $\epsilon_k$ and $\eta_k$ are correlated, we need to construct the joint distribution over $X_k$ and its generating noise $\epsilon_k$ and form the augmented two-dimensional state 
$\zeta_k = 
\begin{bmatrix}
	X_k & \epsilon_{k}
\end{bmatrix}^\top$. 
The dynamics of $\zeta_k$ are then given by
\begin{equation}\label{eq:stacked-vol}
	\begin{pmatrix}
		X_{k+1} \\ \epsilon_{k+1}
	\end{pmatrix} = 
	\begin{pmatrix}
		\alpha & \sigma \\ 
		0 & 0
	\end{pmatrix}
	\begin{pmatrix}
		X_{k} \\ \epsilon_{k}
	\end{pmatrix}
	+ 
	\begin{pmatrix}
		\mu(1 - \alpha)\\
		0
	\end{pmatrix}
	+ q_{k} \begin{pmatrix}
		0\\
		1
	\end{pmatrix},
\end{equation}
where $q_{k}$ is a standard Gaussian random variable. With these dynamics, we have the following observation model
\begin{equation}
	Y_k = \exp(X_k / 2) (\rho \epsilon_k + \sqrt{1 - \rho^2} \, r_k),
\end{equation}
with $r_k$ being a standard Gaussian random variable. Contrary to the original form of the model, the noise processes are now de-correlated so that we can apply our method to the 2-dimensional system defined by $\zeta_k$.

To perform an empirical comparison, we simulate the model and record three trajectories with $K=1\,000$, $K=1\,500$, and $K=2\,000$ observations. We use parameters $\alpha_*=0.975$, $\mu_* = 0.5$, $\sigma_*^2 = 0.02$, and $\rho_*=-0.8$. These values correspond to those in~\cite{malik2011continuous}, typical of a standard stock market. 

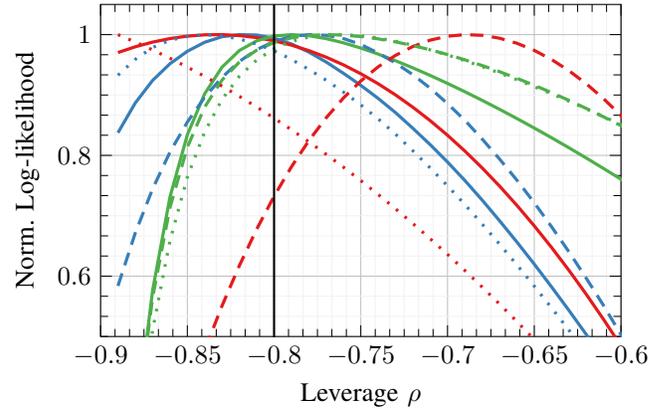
\begin{figure}[!t]
	\centering
\begin{tikzpicture}

\definecolor{crimson2282628}{RGB}{228,26,28}
\definecolor{darkgray176}{RGB}{176,176,176}
\definecolor{mediumseagreen7717574}{RGB}{77,175,74}
\definecolor{steelblue55126184}{RGB}{55,126,184}

\begin{axis}[
width=8.5cm,
height=6cm,
grid=both,
grid style={line width=.1pt, draw=gray!10},
major grid style={line width=.2pt,draw=gray!50},
minor tick num=5,
xmin=-0.9, xmax=-0.6,
xtick style={color=black},
ymin=0.5, ymax=1.05,
ytick style={color=black},
xlabel={Leverage $\rho$},
ylabel={Norm. Log-likelihood}
]
\addplot [very thick, steelblue55126184]
table {%
	-0.89 0.837455918844733
	-0.88 0.88685463725036
	-0.87 0.924222316096537
	-0.86 0.951797812814685
	-0.85 0.972403331078712
	-0.84 0.987366263673575
	-0.83 0.99653385622523
	-0.82 1
	-0.81 0.999395234129435
	-0.8 0.99387811231486
	-0.79 0.984613287448572
	-0.78 0.97114877071728
	-0.77 0.954686473776724
	-0.76 0.935955618573023
	-0.75 0.916010114350214
	-0.74 0.894391996644376
	-0.73 0.870873843288893
	-0.72 0.844897685421829
	-0.71 0.817761953790794
	-0.7 0.788935172855657
	-0.69 0.758345326520088
	-0.68 0.725808385014866
	-0.67 0.6917160802075
	-0.66 0.656443106188792
	-0.65 0.619904012078933
	-0.64 0.582067774786657
	-0.63 0.543186719304072
	-0.62 0.503243927097533
	-0.61 0.462182403348435
	-0.6 0.420032622447984
	-0.59 0.376833005499214
	-0.58 0.332541250790647
	-0.57 0.287331998087183
	-0.56 0.241192840325345
	-0.55 0.194649794688399
	-0.54 0.14701614115124
	-0.53 0.0988480835855815
	-0.52 0.0498273649934876
	-0.51 0
};
\addplot [very thick, crimson2282628]
table {%
	-0.89 0.970306779622165
	-0.88 0.979878547277191
	-0.87 0.987604621729665
	-0.86 0.993482602003073
	-0.85 0.997509818888247
	-0.84 0.999683344279185
	-0.83 1
	-0.82 0.998456366159872
	-0.81 0.995048789054683
	-0.8 0.989773388634346
	-0.79 0.982626065563853
	-0.78 0.973602507882675
	-0.77 0.962698197290029
	-0.76 0.94990841507335
	-0.75 0.935228247678226
	-0.74 0.918652591954208
	-0.73 0.900176160073729
	-0.72 0.879793484141074
	-0.71 0.857498920505231
	-0.7 0.833286653782109
	-0.69 0.807150700595988
	-0.68 0.779084913057034
	-0.67 0.749082981968703
	-0.66 0.717138439794908
	-0.65 0.68324466336883
	-0.64 0.6473948763732
	-0.63 0.609582151589611
	-0.62 0.569799412919582
	-0.61 0.528039437197285
	-0.6 0.484294855785444
	-0.59 0.438558155970112
	-0.58 0.390821682159786
	-0.57 0.341077636889142
	-0.56 0.289318081639581
	-0.55 0.235534937479747
	-0.54 0.179719985529586
	-0.53 0.12186486725817
	-0.52 0.0619610846133424
	-0.51 0
};
\addplot [very thick, mediumseagreen7717574]
table {%
	-0.89 0
	-0.88 0.345728080768015
	-0.87 0.576657482647135
	-0.86 0.731111473452864
	-0.85 0.834086258156238
	-0.84 0.902094145123398
	-0.83 0.946157035364072
	-0.82 0.97369544407862
	-0.81 0.989740896539276
	-0.8 0.997725125842932
	-0.79 1
	-0.78 0.998183608558532
	-0.77 0.993392695687838
	-0.76 0.98639996148018
	-0.75 0.977741208056846
	-0.74 0.967788714224164
	-0.73 0.956801699807412
	-0.72 0.94496114868849
	-0.71 0.932393896872146
	-0.7 0.919189321630338
	-0.69 0.905410914187806
	-0.68 0.891104305561011
	-0.67 0.876302829324327
	-0.66 0.861031372006853
	-0.65 0.845309032356355
	-0.64 0.829150952013852
	-0.63 0.81256957012354
	-0.62 0.795575477886185
	-0.61 0.778177995830232
	-0.6 0.760385559430068
	-0.59 0.742205972814904
	-0.58 0.723646572197756
	-0.57 0.704714328016487
	-0.56 0.685415905930229
	-0.55 0.665757700622922
	-0.54 0.645745852050632
	-0.53 0.625386250719002
	-0.52 0.604684536480517
	-0.51 0.583646093849529
};
\addplot [very thick, steelblue55126184, dash pattern=on 5pt off 3.2pt]
table {%
	-0.89 0.583843094777158
	-0.88 0.671558497538978
	-0.87 0.744612526106411
	-0.86 0.805457645215348
	-0.85 0.855121985376604
	-0.84 0.896365821141741
	-0.83 0.929211758419368
	-0.82 0.955642219115378
	-0.81 0.975094613691008
	-0.8 0.988834442736796
	-0.79 0.996999213361292
	-0.78 1
	-0.77 0.998943084512402
	-0.76 0.993476753121704
	-0.75 0.98373604269857
	-0.74 0.969537269625882
	-0.73 0.952980576025659
	-0.72 0.933699650650782
	-0.71 0.911305795514322
	-0.7 0.885646376025655
	-0.69 0.858074223332307
	-0.68 0.826907704010997
	-0.67 0.793900974297415
	-0.66 0.758915326907413
	-0.65 0.721545646769706
	-0.64 0.681631548343071
	-0.63 0.640187712144537
	-0.62 0.596404444874743
	-0.61 0.55106279594572
	-0.6 0.502891649961062
	-0.59 0.453079970547827
	-0.58 0.401773694611129
	-0.57 0.348920747824032
	-0.56 0.294638782458921
	-0.55 0.238297096380262
	-0.54 0.180461019166851
	-0.53 0.121144757051241
	-0.52 0.0615537464635727
	-0.51 0
};
\addplot [very thick, crimson2282628, dash pattern=on 5pt off 3.2pt]
table {%
	-0.89 0
	-0.88 0.11168094725542
	-0.87 0.214854829168217
	-0.86 0.309961570643602
	-0.85 0.397411262318455
	-0.84 0.477586048657592
	-0.83 0.550841928868529
	-0.82 0.61751046552043
	-0.81 0.677900398567925
	-0.8 0.732299164572973
	-0.79 0.780974322477776
	-0.78 0.824174888345267
	-0.77 0.862132582318737
	-0.76 0.895062991522125
	-0.75 0.923166652984705
	-0.74 0.946630060820931
	-0.73 0.965626601987218
	-0.72 0.980317424883182
	-0.71 0.990852245000155
	-0.7 0.997370091699227
	-0.69 1
	-0.68 0.998861651130179
	-0.67 0.994065965359327
	-0.66 0.985715650443034
	-0.65 0.973905708823672
	-0.64 0.958723906505497
	-0.63 0.940251206347986
	-0.62 0.918562168331189
	-0.61 0.893725319147187
	-0.6 0.865803493334974
	-0.59 0.834854147972265
	-0.58 0.800929652818042
	-0.57 0.764077557635882
	-0.56 0.724340838299849
	-0.55 0.681758123152339
	-0.54 0.636363900971381
	-0.53 0.588188711790105
	-0.52 0.537259321704811
	-0.51 0.483598882724974
};
\addplot [very thick, mediumseagreen7717574, dash pattern=on 5pt off 3.2pt]
table {%
	-0.89 0
	-0.88 0.340230071387239
	-0.87 0.564728133834963
	-0.86 0.713828343976137
	-0.85 0.813360639457268
	-0.84 0.880031602151345
	-0.83 0.924727012101797
	-0.82 0.954579902769451
	-0.81 0.974285099555326
	-0.8 0.986944202225052
	-0.79 0.99461334489595
	-0.78 0.998660554128289
	-0.77 1
	-0.76 0.999246075410004
	-0.75 0.996814974457489
	-0.74 0.992991747670814
	-0.73 0.9879745892241
	-0.72 0.981904080719834
	-0.71 0.974882486229516
	-0.7 0.96698646735927
	-0.69 0.958275449296133
	-0.68 0.948797116291191
	-0.67 0.938591016232836
	-0.66 0.927690922901462
	-0.65 0.916126384670078
	-0.64 0.903923742483913
	-0.63 0.891106803122269
	-0.62 0.877697289642272
	-0.61 0.863715148524243
	-0.6 0.849178765073566
	-0.59 0.834105120340932
	-0.58 0.818509910758661
	-0.57 0.802407643959886
	-0.56 0.785811719142704
	-0.55 0.768734497183174
	-0.54 0.751187363614364
	-0.53 0.733180786338038
	-0.52 0.714724369153582
	-0.51 0.695826901719699
};
\addplot [very thick, steelblue55126184, dash pattern=on 1pt off 3.3pt]
table {%
	-0.89 0.932767993545534
	-0.88 0.959091361063926
	-0.87 0.978732962336865
	-0.86 0.990752711906315
	-0.85 0.996947607783172
	-0.84 1
	-0.83 0.997987647522562
	-0.82 0.992738271359382
	-0.81 0.984423169024218
	-0.8 0.973136437046365
	-0.79 0.959006728627441
	-0.78 0.942419149060017
	-0.77 0.924631825633397
	-0.76 0.904780629967516
	-0.75 0.883692389108221
	-0.74 0.860468791459502
	-0.73 0.835157608799229
	-0.72 0.807617301752428
	-0.71 0.779196677747882
	-0.7 0.749447241538778
	-0.69 0.718608302994307
	-0.68 0.686108270016134
	-0.67 0.653023875050261
	-0.66 0.618632291159602
	-0.65 0.583347224575428
	-0.64 0.547333067747802
	-0.63 0.510499464762952
	-0.62 0.471891167780054
	-0.61 0.432156430243417
	-0.6 0.392055674856619
	-0.59 0.351626152682972
	-0.58 0.310109351595639
	-0.57 0.26823090938411
	-0.56 0.225419405557898
	-0.55 0.181568685796885
	-0.54 0.136583402792451
	-0.53 0.0916224155935902
	-0.52 0.0460740254200012
	-0.51 0
};
\addplot [very thick, crimson2282628, dash pattern=on 1pt off 3.3pt]
table {%
	-0.89 1
	-0.88 0.987710593752688
	-0.87 0.974644905688917
	-0.86 0.960805986153454
	-0.85 0.946196685245758
	-0.84 0.930819657266984
	-0.83 0.914677364965998
	-0.82 0.897772083586188
	-0.81 0.880105904728444
	-0.8 0.861680740039044
	-0.79 0.842498324727292
	-0.78 0.822560220925496
	-0.77 0.801867820897407
	-0.76 0.780422350099355
	-0.75 0.758224870105413
	-0.74 0.735276281401831
	-0.73 0.711577326054641
	-0.72 0.687128590258154
	-0.71 0.661930506770649
	-0.7 0.635983357240455
	-0.69 0.609287274428301
	-0.68 0.581842244331498
	-0.67 0.553648108213533
	-0.66 0.524704564542293
	-0.65 0.49501117084395
	-0.64 0.464567345473905
	-0.63 0.433372369307985
	-0.62 0.401425387361709
	-0.61 0.368725410333729
	-0.6 0.335271316085438
	-0.59 0.301061851050896
	-0.58 0.266095631588263
	-0.57 0.230371145268436
	-0.56 0.19388675210955
	-0.55 0.156640685756011
	-0.54 0.118631054605759
	-0.53 0.0798558428902677
	-0.52 0.0403129117051405
	-0.51 0
};
\addplot [very thick, mediumseagreen7717574, dash pattern=on 1pt off 3.3pt]
table {%
	-0.89 0
	-0.88 0.300978025618126
	-0.87 0.510806479343413
	-0.86 0.6587274763676
	-0.85 0.763884209208533
	-0.84 0.839030443881436
	-0.83 0.892795601266815
	-0.82 0.931101218097082
	-0.81 0.95806481125373
	-0.8 0.976586472235929
	-0.79 0.98873496055856
	-0.78 0.996004884611921
	-0.77 0.999489823107163
	-0.76 1
	-0.75 0.998143033144772
	-0.74 0.994379884215474
	-0.73 0.989064025389576
	-0.72 0.982469160705231
	-0.71 0.974809079652618
	-0.7 0.966252054587347
	-0.69 0.956931416712761
	-0.68 0.946953425497705
	-0.67 0.936403197144565
	-0.66 0.9253492223096
	-0.65 0.913846844012345
	-0.64 0.901940958416904
	-0.63 0.889668127121924
	-0.62 0.877058238498368
	-0.61 0.864135819959336
	-0.6 0.850921077803377
	-0.59 0.83743072309774
	-0.58 0.823678628756356
	-0.57 0.80967635301985
	-0.56 0.795433557009446
	-0.55 0.78095833819427
	-0.54 0.76625749708791
	-0.53 0.751336750907538
	-0.52 0.736200905101389
	-0.51 0.720853991404317
};
\addplot [thick, black]
table {%
	-0.8 -0.05
	-0.8 1.05
};
\end{axis}

\end{tikzpicture}
	\vspace{-0.75cm}
	\caption{Comparing the (normalized) marginal log-likelihood as a function of the leverage parameter $\rho$ in a stochastic volatility model. We plot the estimates as returned by an extended Kalman filter (red), a bootstrap particle filter (blue), and a variational Wasserstein filter (green). The dotted, dashed, and solid lines correspond to different numbers of observations $K=1\,000$, $K=1\,500$, and $K=2\,000$. The true value is $\rho=-0.8$ (vertical line). The variational Wasserstein and particle filters deliver consistent approximations independent of the data size, while the extended Kalman filter does not.}
	\label{fig:sv_lklhd_exp}
\end{figure}

To illustrate the problem of using methods based on linearizing the conditional observation mean $\mathbb{E}\left[Y_k \given X_k, \epsilon_k\right]$, we evaluate the marginal log-likelihood of the data approximated by an extended Kalman filter (EKF) targeting \eqref{eq:stacked-vol} with parameters $(\alpha_*, \mu_*, \sigma_*)$ while varying levels of correlation $\rho$ and number of observations $K$. The resulting curves for the particle filter of~\cite{malik2011continuous}, our method, and the extended Kalman filter are shown in Figure~\ref{fig:sv_lklhd_exp}. As the (model) correlation $\rho$ between the two noise-generating processes decreases, the predictive value of the observations becomes negligible. As a consequence, the marginal likelihood, as approximated by the EKF, will struggle to capture the right level of correlation. This means that calibrating the model using an extended (or similar) Kalman filter will result in an inconsistent estimate, at least for the parameter $\rho$. 

To further confirm this hypothesis, we follow \cite{malik2011continuous} and perform joint maximum likelihood parameter estimation given $K=1\,000$ observations. We compare our method to theirs and to an extended Kalman filter. The likelihood is optimized in all cases using the gradient of the log-likelihood, delivered by automatic reverse differentiation. In the particular case of the variational Wasserstein filter, differentiation is made efficient by our fixed point formulation of the variational calibration. 

For a statistical comparison, we repeat this experiment over ten trials associated with different random seeds and, therefore, ten distinct sets of observations. We report the mean plus or minus one standard deviation of the parameter estimates for each algorithm. However, because the particle filter is inherently a stochastic method, even for a fixed set of observations, we perform $25$ \emph{sub-trials} per set of observations and use the median as a parameter estimate per trial. The results are given in Table~\ref{tab:mle-stochvol} and confirm our intuition. The extended Kalman filter provides inconsistent solutions, while our method delivers estimates comparable to the estimates of the particle filter.

\begin{table}[!t]
	\centering
	\caption{Statistics for parameter estimation of the stochastic volatility model. Results are averaged over ten trials.}
	\label{tab:mle-stochvol}
	\begin{adjustbox}{max width=\columnwidth}
		\begin{tabular}{lcccc}
			\toprule
			& $\mu$ & $\alpha$ & $\sigma$ & $\rho$ \\
			\midrule
			True & 0.50 & 0.975 & 0.14 & -0.80 \\
			\midrule
			PF \cite{malik2011continuous} 	& $0.56 \, (\pm 0.08)$ & $0.972 \, (\pm 0.007)$ & $0.15 \, (\pm 0.02)$ & $-0.85 \, (\pm 0.06)$ \\
			VWF 							& $0.56 \, (\pm 0.07)$ & $0.972 \, (\pm 0.009)$ & $0.15 \, (\pm 0.02)$ & $-0.80 \, (\pm 0.04)$ \\
			EKF 							& $0.69 \, (\pm 0.33)$ & $0.780 \, (\pm 0.590)$ & $0.21 \, (\pm 0.25)$ & $-0.58 \, (\pm 0.54)$ \\
			\bottomrule
		\end{tabular}
	\end{adjustbox}
\end{table}

\subsection{Multi-modal Example}
We now turn to the multi-modal motivating example \eqref{eq:multimodal}. We simulate $K=500$ observations from the model, where we have taken the variance at origin to be $\delta^2 = 1$. We then perform variational filtering using the mixture version of Algorithm~\ref{alg:variational-filtering} with $N_i=2$ mixture components. 

\begin{figure}[!t]
	\centering
	\input{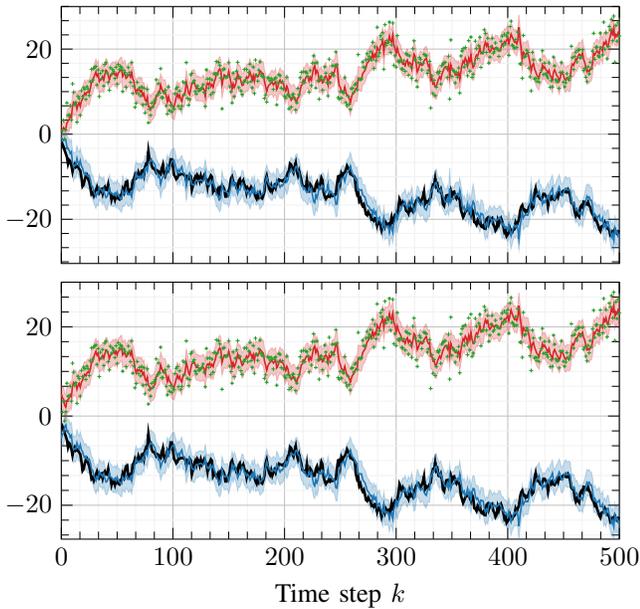}
	\vspace{-0.5cm}
	\caption{Performing filtering on a multi-modal dynamical system. We compare the filtering result of a particle filter (top) with that of a variational Wasserstein filter equipped with a mixture-of-Gaussians posterior representation (bottom). Both filters capture the bi-modal posterior distribution (red and blue) inferred from the observations (green) induced by the true states (black).}
	\label{fig:multi_modal_exp}
\end{figure}

Our goal is a correct representation of the filtering posterior. Thus we will not assess the performance in terms of root mean square error, which is inappropriate for such problems. Instead, we qualitatively report the resulting filtering distributions in Figure~\ref{fig:multi_modal_exp}. There is visually hardly any difference between the particle filtering estimate and our method.

One caveat to this positive result is that the bi-modal variational Wasserstein filter tends to collapse when the two modes are too close to each other. It is still unclear whether this is a feature of the general method or the ODE solver, and further investigations are warranted.

\section{Conclusion}
In this article, we have presented a novel approach for approximate Gaussian filtering, which avoids the reliance on \emph{enabling} assumptions~\cite{garcia2015posterior}, which are usually not amenable to multiplicative noise and not extendable to multi-modal distributions. Several questions remain open: 
\paragraph{Assumptions} We have assumed, for simplicity, that the transition dynamics are governed by affine Gaussian densities. Overcoming this limitation is an interesting research question. In fact, in \cite{sarkka2007unscented}, the coupled ODEs~\eqref{eq:mean-cov-flow} were originally introduced to propagate the Gaussian through a nonlinear stochastic differential equation.
\paragraph{Numerics} The numerical scheme chosen here is, primarily for ease of exposition, different from that of~\cite{Lambert2022}, which uses an iterative method called JKO~\cite{santambrogio2017gradientflows} after~\cite{jordan1998variational}, instead of integrating the differential equation~\eqref{eq:mean-cov-flow} directly. It is unclear which approach is the best fit for filtering applications.
\paragraph{Mixture weights} The weights of the mixture in the ODE~\eqref{eq:mean-cov-flow-approx} are not allowed to vary. This modeling restriction needs to be relaxed. However, that may lead to identifiability issues (for example, a single Gaussian can be represented with a mixture of two Gaussians with different weights in infinite ways). Furthermore, introducing time-varying weights should be done carefully.

\section{Individual Contributions}
The original idea and redaction of the article are due to Adrien Corenflos. Both authors contributed to the design of the methodology. The implementation and experiments are due to Hany Abulsamad.

\bibliographystyle{IEEEtran}
\bibliography{main}

\begin{thebibliography}{10}
\providecommand{\url}[1]{#1}
\csname url@samestyle\endcsname
\providecommand{\newblock}{\relax}
\providecommand{\bibinfo}[2]{#2}
\providecommand{\BIBentrySTDinterwordspacing}{\spaceskip=0pt\relax}
\providecommand{\BIBentryALTinterwordstretchfactor}{4}
\providecommand{\BIBentryALTinterwordspacing}{\spaceskip=\fontdimen2\font plus
\BIBentryALTinterwordstretchfactor\fontdimen3\font minus
  \fontdimen4\font\relax}
\providecommand{\BIBforeignlanguage}[2]{{%
\expandafter\ifx\csname l@#1\endcsname\relax
\typeout{** WARNING: IEEEtran.bst: No hyphenation pattern has been}%
\typeout{** loaded for the language `#1'. Using the pattern for}%
\typeout{** the default language instead.}%
\else
\language=\csname l@#1\endcsname
\fi
#2}}
\providecommand{\BIBdecl}{\relax}
\BIBdecl

\bibitem{jazwinski2007stochastic}
A.~H. Jazwinski, \emph{Stochastic Processes and Filtering Theory}.\hskip 1em
  plus 0.5em minus 0.4em\relax Courier Corporation, 2007.

\bibitem{Kalman1960}
R.~E. Kalman, ``A new approach to linear filtering and prediction problems,''
  \emph{Journal of Basic Engineering}, 1960.

\bibitem{sarkka2013bayesian}
S.~S{\"a}rkk{\"a}, \emph{{B}ayesian Filtering and Smoothing}.\hskip 1em plus
  0.5em minus 0.4em\relax Cambridge University Press, 2013.

\bibitem{Chopin2020book}
N.~Chopin and O.~Papaspiliopoulos, \emph{An Introduction to Sequential {M}onte
  {C}arlo}.\hskip 1em plus 0.5em minus 0.4em\relax Springer, 2020.

\bibitem{heston1993closed}
S.~L. Heston, ``A closed-form solution for options with stochastic volatility
  with applications to bond and currency options,'' \emph{The Review of
  Financial Studies}, 1993.

\bibitem{garcia2015posterior}
A.~F. Garc\`ia-Fern\`andez, L.~Svensson, M.~R. Morelande, and S.~S\"arkk\"a,
  ``Posterior linearization filter: {P}rinciples and implementation using sigma
  points,'' \emph{{IEEE} Transactions on Signal Processing}, 2015.

\bibitem{Tronarp2018iterative}
F.~Tronarp, {\'A}.~F. Garc\'ia-Fern\'andez, and S.~S\"arkk\"a, ``Iterative
  filtering and smoothing in nonlinear and non-{G}aussian systems using
  conditional moments,'' \emph{{IEEE} Signal Processing Letters}, 2018.

\bibitem{Lambert2022}
M.~Lambert, S.~Chewi, F.~Bach, S.~Bonnabel, and P.~Rigollet, ``Variational
  inference via {Wasserstein} gradient flows,'' \emph{arXiv preprint
  arXiv:2205.15902}, 2022.

\bibitem{joyce2011kullback}
J.~M. Joyce, ``{Kullback--Leibler} divergence,'' in \emph{International
  Encyclopedia of Statistical Science}.\hskip 1em plus 0.5em minus 0.4em\relax
  Springer, 2011.

\bibitem{murphy2022book}
K.~P. Murphy, \emph{Probabilistic Machine Learning: {A}n Introduction}.\hskip
  1em plus 0.5em minus 0.4em\relax MIT Press, 2022.

\bibitem{santambrogio2017gradientflows}
F.~Santambrogio, ``{Euclidean, metric, and Wasserstein} gradient flows: {A}n
  overview,'' \emph{Bulletin of Mathematical Sciences}, 2017.

\bibitem{Villani2009}
C.~Villani, \emph{Optimal Transport: {O}ld and New}.\hskip 1em plus 0.5em minus
  0.4em\relax Springer Berlin Heidelberg, 2009.

\bibitem{sarkka2007unscented}
S.~S\"arkk\"a, ``On unscented {K}alman filtering for state estimation of
  continuous-time nonlinear systems,'' \emph{{IEEE} Transactions on Automatic
  Control}, 2007.

\bibitem{pages2018numerical}
G.~Pag{\`e}s, \emph{Numerical Probability}.\hskip 1em plus 0.5em minus
  0.4em\relax Springer, 2018.

\bibitem{christianson1994fixedpoint}
B.~Christianson, ``Reverse accumulation and attractive fixed points,''
  \emph{Optimization Methods \& Software}, 1994.

\bibitem{gordon1993novel}
N.~J. Gordon, D.~J. Salmond, and A.~F. Smith, ``Novel approach to
  nonlinear/non-{Gaussian Bayesian} state estimation,'' in \emph{IEEE
  Proceedings of Radar and Signal Processing}, 1993.

\bibitem{malik2011continuous}
S.~Malik and M.~K. Pitt, ``Particle filters for continuous likelihood
  evaluation and maximisation,'' \emph{Journal of Econometrics}, 2011.

\bibitem{jordan1998variational}
R.~Jordan, D.~Kinderlehrer, and F.~Otto, ``The variational formulation of the
  {Fokker--Planck} equation,'' \emph{SIAM Journal on Mathematical Analysis},
  1998.

\end{thebibliography}
	
\end{document}